\documentstyle[psfig,preprint,aps]{revtex}
\begin{document}
\draft
\preprint{MKPH-T-97-15}
\title{
{\bf 
Influence of final state interaction on incoherent $\eta$-photoproduction 
on the deuteron near threshold 
}
\footnote{Supported by the Deutsche Forschungsgemeinschaft (SFB 201)}}

\author{A.\ Fix\footnote
{Supported by the Deutscher Akademischer Austauschdienst} 
and H.\ Arenh\"ovel}
\address{
Institut f\"{u}r Kernphysik, Johannes Gutenberg-Universit\"{a}t,
       D-55099 Mainz, Germany}
\date{\today}
\maketitle

\begin{abstract}
Incoherent $\eta$-photoproduction on the deuteron is studied for 
photon energies from threshold up to 720 MeV. 
The elementary $\gamma N\!\rightarrow\!\eta N$ amplitude 
is constructed in the frame of a conventional isobar model. 
Effects of final state interaction are investigated 
and their role in the cross section as well as in polarization 
observables is found to be important. Inclusion of such effects leads to 
a satisfactory agreement with experimental data in the near threshold region.
In addition, recent experimental results on the coherent reaction 
$d(\gamma,\eta)d$ are discussed.
\end{abstract}

\section{Introduction}
The main motivation for studying $\eta$-photoproduction on the deuteron
is to obtain information on the elementary process on the neutron.
The evident advantages of using a deuteron target are 
(i) the small binding energy of nucleons in the deuteron, which  
from the kinematical point of view provides the case of a nearly free 
neutron target, and (ii) minimal number of nucleons, reducing the influence 
of a nuclear environment on the elementary production process and allowing
a careful microscopic treatment. 
The interest in the elementary process on the neutron is closely 
connected with the question concerning the isotopic structure of 
the $S_{11}$(1535) photoexcitation amplitude. A combined analysis of 
$\gamma p\!\rightarrow\!\eta p$, 
$\gamma d\!\rightarrow\!\eta d\,$ and 
inclusive $\gamma d\!\rightarrow\!\eta X$ reactions 
\cite{Kru95,Bacci69} have confirmed its predominantly 
isovector character as predicted by quark models \cite{Konjuk80}. 
Furthermore, very recently exclusive measurements of the ($\gamma,\eta N$)
reaction on the deuteron for quasifree kinematics with detection 
of forward emerging nucleons in coincidence 
with etas were performed \cite{Hoff97}. This experiment, 
providing direct information on the ratio $\sigma_n/\sigma_p$ ,
is expected to be quite precise, since the experimental systematic 
errors as well as different model corrections will largely cancel. 
From the data of \cite{Hoff97} the ratio 
$\alpha=t^{(s)}_{\gamma\eta}/t^{(p)}_{\gamma\eta}$
of the isoscalar to the proton amplitude is extracted as 
$\alpha_{qf}=0.09\pm0.02$. At the same time, if one wants to find 
agreement between the data on the coherent $\gamma d\!\rightarrow\!\eta d$
cross section of \cite{Hoff97} and conventional 
impulse approximation 
calculations \cite{Breit97,Kamalov97}, one needs a considerably larger 
parameter $\alpha_{coh}=0.20\pm0.02$. 
If one assumes that the value $\alpha_{qf}$ is more reliable, one is lead  
to the conclusion that the impulse approximation fails 
in describing the coherent reaction. Thus one either would need
to search for new mechanisms dominating the coherent process 
or one would need to question the accuracy of the experimental data.

It should be noted, that the method of extracting the information 
about the single nucleon processes from the deuteron data is based on the 
so-called spectator-nucleon model, in
which the pure quasifree production is considered 
as the only mechanism for the knock-out $d(\gamma,\eta N)N$ reaction.
Therefore, a careful investigation of the validity of this approximation,  
i.e., the estimation of possible other effects is needed. The main goal of 
the present work is to study the role of rescattering in the final state 
of the $\gamma d\!\rightarrow\!\eta np$ process. One can expect that 
such effects become 
important close to the reaction threshold. 
In this region the 
smallness of the excitation energy in the final $np$-system and the large 
momentum transfer (which is about the $\eta$ mass in the 
$\gamma d$ c.m.\ frame) lead to a kinematical 
situation, where two final nucleons move primarily 
together with a large total, but small relative momentum. 
For this kinematics, the spectator model is expected to give 
a very small cross section since the momenta of both nucleons are large
and, on the other hand, the corrections due to the strong 
$NN$-interaction may be significant. Furthermore, as has been shown in
\cite{Fix97}, the $\eta$-rescattering can also visibly change the 
$\gamma d\!\rightarrow\!\eta np$ cross section near threshold.
 
In Section \ref{section2} we present the isobar model for the elementary 
$\eta$-photoproduction process. The treatment of the 
$\gamma d\!\rightarrow\!\eta np$ amplitude,
based on time-ordered perturbation theory including $\eta N$
and $NN$-rescattering, is developed in Section \ref{section3}. 
A similar approach was adopted for the analysis of coherent 
$\pi^0$-photoproduction on the deuteron \cite{Paul95}. In our notation, we 
follow mainly this paper.   
In this Section, we also discuss the effects of rescattering on cross 
sections and polarization observables and compare our results with 
available experimental data. 
Finally, some attention is paid to the interpretation of 
experimental results of 
Ref.\ \cite{Hoff97} on the coherent process. 

\section{The $\gamma N\!\rightarrow\!\eta N$ amplitude}\label{section2}

As a starting point we consider the elementary process 
\begin{equation}\label{11}
\gamma(k_{\mu},\vec{\varepsilon}_\lambda)+N(p_{\mu})\rightarrow\eta(q_{\mu})
+N(p_{\mu}^{\,\prime})\,,
\end{equation}
which we need as input for studying the $\eta$-photoproduction 
on the deuteron. 
The four momenta of the participating particles are denoted by  
\begin{equation}\label{12}
k_{\mu}=(\omega,\vec{k}\,),\quad  q_{\mu}=(\omega_{\vec{q}},\vec{q}\,),\quad  
p_{\mu}=(E_{\vec{p}},\vec{p}\,),\quad p_{\mu}^{\,\prime}
=(E_{\vec{p}^{\,\prime}},
\vec{p}^{\,\prime}) \,,
\end{equation}
and $\vec{\varepsilon}_\lambda$ stands for the photon polarization vector. 
The particle energies are 
\begin{equation}\label{13}
\omega=k,\quad   \omega_{\vec{q}}=\sqrt{q^2+m_{\eta}^2},
\quad    E_{\vec{p}^{\,(\prime)}}=\sqrt{p^{\,(\prime)2}+M_N^2}\,,
\end{equation}
where $M_N$ and $m_{\eta}$ are the masses of nucleon and $\eta$ meson, 
and $k$, $q$, $p$, and $p^{\,\prime}$ denote the absolute values of the 
respective three-momenta.

Past studies \cite{Kru95a,Mukh95,Tiat94,Sau95} of the 
$p(\gamma,\eta)p$ process for photon lab energies from threshold
up to 800 MeV have clearly shown the dominant role of the 
$S_{11}$(1535) resonance
which decays with a relatively large probability into the $\eta N$ channel.
For this reason we have chosen a nonrelativistic isobar model, 
in which the $\eta$-photoproduction proceeds exclusively  
through the electric dipole transition to the $S_{11}$(1535) (hereafter
denoted by $N^*$). Possible 
background contributions, such as nucleon pole and vector meson exchange,
are neglected because of their insignificance in the energy region
under consideration. Within the mentioned model the elementary 
$\eta$-photoproduction amplitude in an arbitrary frame of reference reads    
\begin{equation}\label{14}
t_{\gamma\eta}(W_{\gamma N})
=v_{\eta N^*}^{\dagger}g_{N^*}(W_{\gamma N})v_{\gamma N^*}\,,
\end{equation}
where $W_{\gamma N}$ is the $\gamma N$ invariant mass.
The nonrelativistic current for the electromagnetic vertex
$v_{\gamma N^*}\!=\!-\vec{\varepsilon}_{\lambda}
\cdot\vec{j}_{N^*N}$ is given by
\begin{equation}\label{15}
\vec{j}_{N^*N}=e\frac{k_{\gamma N}}{M_{N^*}+M_N}g_{\gamma NN^*}\vec{\sigma},
\end{equation}\label{16} where $M_{N^*}$ 
is the resonance mass and $k_{\gamma N}$ stands for the relative
photon-nucleon momentum 
\begin{equation}\label{17}
\vec{k}_{\gamma N}=\frac{M_N\vec{k}-(M_{N^*}-M_N)\,\vec{p}}{M_{N^*}}\,.
\end{equation}
The vertex constant $g_{\gamma NN^*}$ can be expressed in terms of  
the helicity amplitude $A_{1/2}^N$ of the $N^*$ photoexcitation 
\begin{equation}\label{18}
e\,g_{\gamma NN^*}=\sqrt{\frac{2M_{N^*}(M_{N^*}\!+\!M_N)}
{M_{N^*}-M_N}}A_{1/2}^N\,.
\end{equation}
Furthermore, we introduce the isoscalar $g_{\gamma NN^*}^{(s)}$ 
and isovector $g_{\gamma NN^*}^{(v)}$ coupling constants 
in accordance with the 
isotopic structure of the photoproduction amplitude for an isoscalar meson 
\begin{equation}\label{19}
g_{\gamma NN^*}=g_{\gamma NN^*}^{(s)}+\tau_3g_{\gamma NN^*}^{(v)}\,.
\end{equation}
For the hadronic vertex $v^{\dagger}_{\eta N^*}$ we use the 
pseudoscalar coupling
\begin{equation}\label{20}
v_{\eta N^*}^{\dagger}=-ig_{\eta NN^*}\,.
\end{equation}
Finally, the $N^*$ propagator 
\begin{equation}\label{21}
g_{N^*}(W_{\gamma N})
=\frac{1}{W_{\gamma N}-M_{N^*}+\frac{i}{2}\Gamma(W_{\gamma N})}
\end{equation}
involves the energy dependent decay width  
\begin{equation}\label{22}
\Gamma=\Gamma_{\eta N}+\Gamma_{\pi N}+\Gamma_{\pi\pi N}\,.
\end{equation}
The partial $\eta N$ and $\pi N$ widths are related to the corresponding 
coupling constants in our nonrelativistic approach 
\begin{equation}\label{23}
\Gamma_{xN}(W_{\gamma N})=\frac{g_{xNN^*}^2}{4\pi}\frac{2M_N}{W_{\gamma N}}
q_x^*,\,\,  (x=\eta,\pi) \,,
\end{equation}
where  
$q_x^*$ is the momentum of the respective particle in the $\gamma N$ 
c.m.\ system.
Following Ref.\ \cite{Breit97} we treat the $N^*$ width for the
$\pi\pi N$
channel purely phenomenologically, as a constant above $\pi\pi N$ threshold.

\begin{figure}
\centerline{\psfig{figure=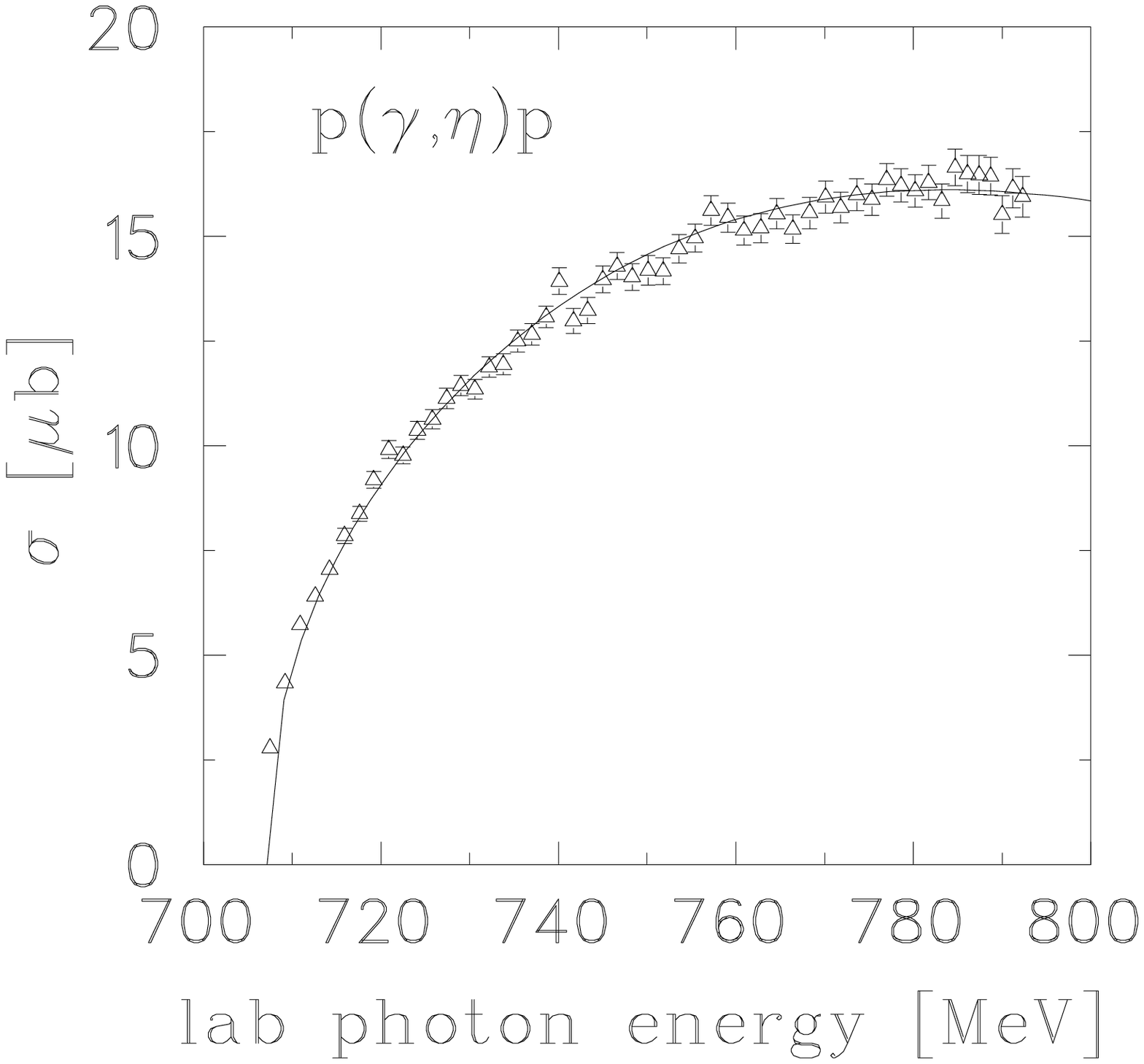,width=8cm,angle=0}}
\vspace*{.5cm}
\caption{
The total cross section of the $p(\gamma,\eta)p$ reaction. The 
experimental points are from \protect{\\} Ref.\ \protect{\cite{Kru95}}.
}
\label{fig1}
\end{figure}

Thus the full $\eta$-photoproduction amplitude may be written in the form
\begin{equation}\label{25}
t_{\gamma\eta}^{(s/v)}=eg_{\eta NN^*}\frac{k_{\gamma N}}
{M_{N^*}+M_N}\,\,\frac{g_{\gamma NN^*}^{(s/v)}}
{W_{\gamma N}-M_{N^*}+\frac{i}{2}
\Gamma(W_{\gamma N})}i\vec{\sigma}\cdot\vec{\varepsilon}_{\lambda} \,.
\end{equation}
In the actual calculation  we use the following set of $N^*$ parameters
\begin{equation}\label{26}
M_{N^*}=1535\,\mbox{MeV},\quad    g_{\eta NN^*}=2.10, 
\quad   g_{\pi NN^*}=1.19,\quad   \Gamma_{\pi\pi N}=16\,\mbox{MeV} \,,
\end{equation}
which gives for the total and partial decay widths at the resonance
position $W_{\gamma N}=M_{N^*}$
\begin{equation}
\Gamma=160\,\mbox{MeV},\quad  \Gamma_{\eta N}=0.5\,{\Gamma},\quad
\Gamma_{\pi N}=0.4\,{\Gamma},\quad  \Gamma_{\pi\pi N}=0.1\,{\Gamma}.
\end{equation}
This parametrization is reasonably consistent 
with the values following from the 1996 PDG listings 
\cite{PDG96}. The amplitude (\ref{25}) gives quite a good description of the  
$\gamma p\!\rightarrow\!\eta p$ total cross section data \cite{Kru95a}
(see Fig.\ \ref{fig1}) if we take for the proton helicity amplitude the value
$A_{1/2}^p=104\cdot10^{-3}$GeV$^{-1/2}$. 

\section{The $\gamma d\!\rightarrow\!\eta np$ reaction}\label{section3}

Now we turn to incoherent $\eta$-photoproduction on the deuteron
\begin{equation}\label{31}
\gamma(k_{\mu},\vec{\varepsilon}_\lambda)+d(Q_{\mu})\!\rightarrow\!
\eta(q_{\mu})+N_1(p_{1\mu})+N_2(p_{2\mu})\,.
\end{equation}
The momenta of all particles are defined similarly to those in 
Section \ref{section2}.
We restrict ourselves to the small region of incident photon energies from 
threshold up to 720 MeV. 
All calculations refer to the laboratory frame, $Q_{\mu}\!=\!(M_d,\vec{0})$, 
unless otherwise noted. The coordinate system is chosen to have a right-hand
orientation with $z$-axis along 
the photon momentum $\vec{k}$ and $y$-axis parallel to 
$\vec{k}\times\vec{q}$. 
We work in the isospin formalism and consider the 
two outgoing nucleons symmetrically, not specifying their isotopic states.
The ratio $\alpha=t^{(s)}_{\gamma\eta}/t^{(p)}_{\gamma\eta}$
is taken to be 0.11, which, as will be discussed below, 
produces the best agreement with experimental
data. The laboratory cross section for the reaction (\ref{31}) with 
unpolarized particles reads
\begin{equation}\label{32}
d\sigma=\frac{1}{(2\pi)^5}\,\delta^4(k_{\mu}+Q_{\mu}-q_{\mu}-p_{1\mu}-
p_{2\mu})\,\frac{1}{6}\sum_{\lambda m_d}\sum_{Sm_ST}
|M_{Sm_ST,\lambda m_d}|^2\,\frac{M_N^2}{2\omega}
\frac{d^3q}{2\omega_{\vec{q}}}\frac{d^3p_1}{E_{\vec{p}_1}}
\frac{d^3p_2}{E_{\vec{p}_2}} \,,
\end{equation}
where $M_{Sm_ST,\lambda m_d}$ denotes the reaction amplitude
\begin{equation}\label{33}
M_{Sm_ST,\lambda m_d}=
\langle\vec{q},\vec{p}_1\vec{p}_2,Sm_ST\,|M_{\lambda}|
\,m_d\rangle \,.\nonumber
\end{equation}
Here $m_d$ stands for the deuteron spin projection, 
and the final two nucleon state is specified by the nucleon momenta 
$\vec{p}_1$, $\vec{p}_2$ as well as by spin
$Sm_S$ and isospin $T(m_T=0)$ quantum numbers. The corresponding fully 
antisymmetric $NN$-function can formally be written as 
\begin{equation}\label{34}
|\,\vec{p}_1\vec{p}_2,Sm_ST\rangle=\frac{1}{\sqrt{2}}
\bigg(|\vec{p}_1\vec{p}_2,Sm_S\rangle
-(-1)^{S+T}|\vec{p}_2\vec{p}_1,Sm_S\rangle\bigg) \,.
\end{equation}
The deuteron wave function in (\ref{33}) is noncovariantly normalized 
to unity. The Fourier transform of its internal part 
separating the spin part has the familiar form
\begin{eqnarray}\label{35}
\phi_{m_Sm_d}(\vec{p}\,)&=&\langle\vec{p},1m_s|1m_d\rangle \\
&=&(2\pi)^{3/2}\sum_{L=0,2}\sum_{m_L}i^L
u_L(p)(Lm_L\,1m_S|1m_d)\,Y_{Lm_L}(\hat{p})\,.\nonumber
\end{eqnarray}
For the radial functions $u_L(p)$ we use the parametrization 
of the Bonn-potential model (OBEPQ-version) \cite{Mach87}.
The main features of the process (\ref{31}) will be investigated by
considering the partially 
integrated cross sections $d\sigma/dq\,d\Omega_{\vec{q}}$
and $d\sigma/d\Omega_{\vec{q}}$, which are obtained from the fully 
exclusive cross section
\begin{equation}
\frac{d\sigma}{d\Omega_{\vec{p}}\,dq\,d\Omega_{\vec{q}}}={(2\pi)^4}
{\bf K}\,\frac{1}{6}\sum_{\lambda m_d}\sum_{Sm_ST}
|M_{Sm_ST,\lambda m_d}|^2\
\end{equation}
by appropriate integration. Here, the kinematic factor 
\begin{equation}\label{36}
{\bf K}=\frac{1}{(2\pi)^9}
\,\,\frac{p^3q^2M_N^2}{4\omega\omega_{\vec{q}}\,\,
\bigg|E_{\vec{p}_1}(p^2+\frac{1}{2}\vec{P}
\!\cdot\!\vec{p})+E_{\vec{p}_2}(p^2-\frac{1}{2}\vec{P}\!\cdot\!\vec{p})
\bigg|} \,
\end{equation}
is expressed in terms of 
relative $\vec{p}\!=\!(\vec{p}_2\!-\!\vec{p}_1)/2$ and 
total $\vec{P}\!=\!\vec{p}_1\!+\!\vec{p}_2$ 
momenta of the two final nucleons.
We prefer this choice of variables, because in this case 
the kinematic factor does not have any singularity on the boundary of 
the available phase space, when $p\!\rightarrow$0 \cite{Rudig95}.
With respect to polarization observables, we consider in this work only
the tensor asymmetries $T_{2M}$ of the deuteron target, which, as will be 
discussed, is most sensitive to the effects of rescattering. 
For this observable, we use the definition similar to that given in 
Ref.\ \cite{Arenh88} for deuteron photodisintegration
\begin{equation}
T_{2M}\frac{d\sigma}{dq\,d\Omega_{\vec{q}}}=
(2-\delta_{M0})\,\Re eV_{2M},\quad   M=0,1,2\,,
\end{equation}
where
\begin{equation}
V_{2M}=\sqrt{\frac{5}{3}}\sum_{m_d^{\prime}m_d}
\sum_{Sm_ST\lambda}(-1)^{1-m_d^{\prime}}
\left(\begin{array}{ccc}
1 & 1 & 2 \\
m_d & -m_d^{\prime} & -M 
\end{array}\right)
\int {\bf K}\,M_{Sm_ST,\lambda m_d}^*M_{Sm_ST,\lambda m_d^{\prime}}
d\Omega_{\vec{p}}\,.
\end{equation}
Thus, for example, the asymmetry $T_{20}$ is given by the expression 
\begin{equation}\label{T20}
\displaystyle T_{20}=\frac{1}{\sqrt{2}}\frac{\sum_{Sm_ST\lambda}\int
{\bf K}\bigg(|M_{Sm_ST,\lambda 1}|^2+
|M_{Sm_ST,\lambda -1}|^2-2|M_{Sm_ST,\lambda 0}|^2\bigg)d\Omega_{\vec{p}}}
{\sum_{\lambda m_d}\sum_{Sm_ST}\int{\bf K}|M_{Sm_ST,\lambda m_d}|^2
d\Omega_{\vec{p}}}\,.
\end{equation}   
Then, the cross section can be expressed in terms 
of the unpolarized cross section and the tensor target asymmetry as
\begin{equation}
\frac{d\sigma(P_2^d)}{dq\,d\Omega_{\vec{q}}}=\frac{d\sigma}
{dq\,d\Omega_{\vec{q}}}\Bigg[1+P_2^d\sum_{M\geq 0}T_{2M}\cos(M\phi_d)
d_{M0}^2(\theta_d)\Bigg],
\end{equation}
where $P_d^2$ is the degree of deuteron tensor polarization with respect 
to the orientation for which the deuteron density matrix is diagonal 
and which is characterized by the angles $\theta_d$ and $\phi_d$ in the
chosen coordinate system. 

\begin{figure}
\centerline{\psfig{figure=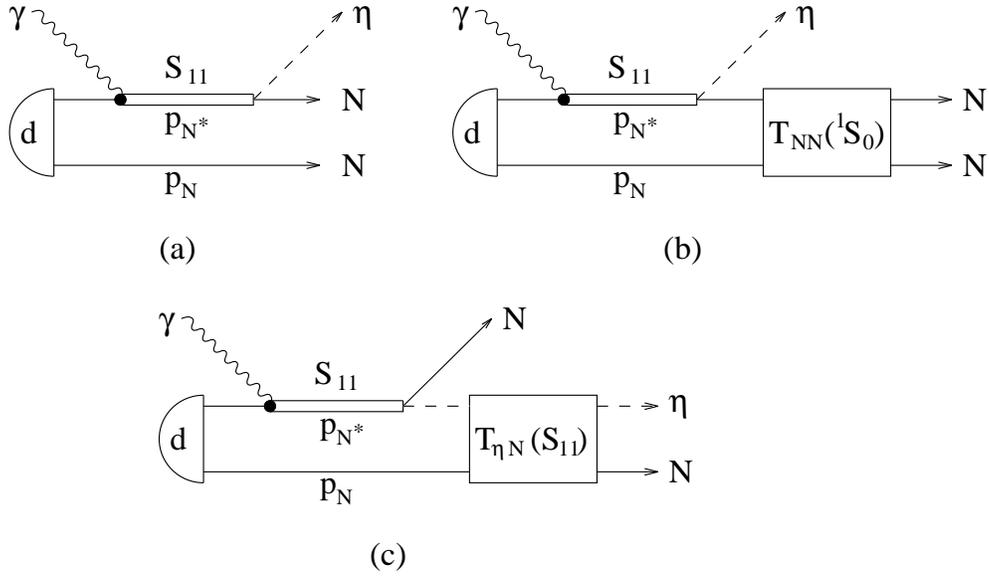,width=13cm,angle=0}}
\vspace*{.5cm}
\caption{
Diagramatic representation of the $\gamma d\!\rightarrow\!\eta np$ amplitude,
as studied in the present work: (a) impulse approximation,
(b) $NN$-rescattering and (c) $\eta N$-rescattering.
}
\label{fig2}
\end{figure}

The transition matrix elements $M_{Sm_ST,\lambda m_d}$ 
are calculated in the frame of 
time-ordered perturbation theory, using the electromagnetic and 
hadronic vertices, introduced in Section \ref{section2}. 
As full amplitude we include besides the pure impulse
approximation (IA) one-loop  
$NN$- and $\eta N$-rescattering terms as shown in Fig.\ \ref{fig2} 
\begin{equation}\label{37}
M_{Sm_ST,\lambda m_d}=M_{Sm_ST,\lambda m_d}^{IA}
+M_{Sm_ST,\lambda m_d}^{NN}+M_{Sm_ST,\lambda m_d}^{\eta N} \,.
\end{equation}
Another possible two-step $\eta$-production mechanism in the 
two-nucleon system $\gamma N\!\rightarrow\!\pi N\!\rightarrow\!\eta N$
gives a negligible contribution as was shown in \cite{Fix97}.
We now will consider successively all three terms and investigate 
their role in the reaction (\ref{31}). 

\subsection{The impulse approximation (IA)}

\begin{figure}
\centerline{\psfig{figure=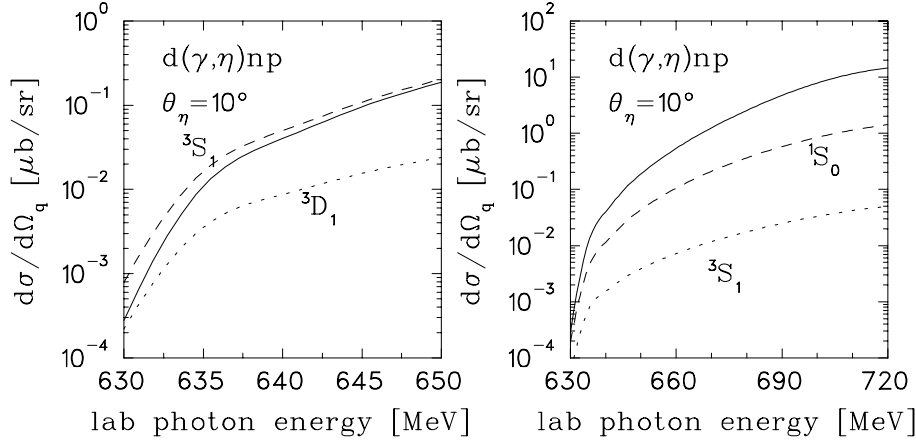,width=12cm,angle=0}}
\vspace*{.5cm}
\caption{
The $\eta$-angular distribution in the $\gamma d\!\rightarrow\!\eta np$
reaction, predicted by the IA (full curve).
The contribution from the deuteron $^3S_1$ and $^3D_1$ components
are separately shown in the left-hand panel. The right-hand panel
presents the separate contributions from the final 
singlet $^1S_0$ and triplet $^3S_1$ waves.   
}
\label{fig3}
\end{figure}

The matrix element given by the diagramm (a) in Fig.\ \ref{fig2} reads
\begin{equation}\label{41}
M_{Sm_ST,\lambda m_d}^{IA}=
-\langle\vec{q},\vec{p}_1\vec{p}_2,
Sm_ST\,|V_{\eta N^*}^{\dagger}\,G_{N^*N}(W_{\gamma N})\,
\vec{\varepsilon}_{\lambda}\cdot\vec{J}_{N^*N}|\,m_d\rangle\,.
\end{equation}
The effective one-particle operators for the 
hadronic vertex $V_{\eta N^*}^{\dagger}$ and 
current $\vec{J}_{N^*N}$ are defined using (\ref{15}) through (\ref{20}) as
\begin{eqnarray}\label{42}
V_{\eta N^*}^{\dagger}&=&v_{\eta N^*}^{\dagger}(1)+
v_{\eta N^*}^{\dagger}(2)\,,\\
\vec{J}_{N^*N}&=&\vec{j}_{N^*N}(1)+\vec{j}_{N^*N}(2)\,,
\end{eqnarray}
with arguments, referring to nucleon 1 and 2. The propagator
for the $N^*N$ intermediate state is
\begin{equation}\label{43}
G_{N^*N}(W_{\gamma N})=\frac{1}{W_{\gamma N}-M_{N^*}+\frac{i}{2}
\Gamma(W_{\gamma N})} \,,
\end{equation}
with $W_{\gamma N}$ being the energy available for the $N^*$ excitation
\begin{equation}\label{44}
W_{\gamma N}=\omega+M_d-M_N-\frac{p_N^2}{2M_N}-\frac{p_{N^*}^2}{2M_{N^*}}\,.
\end{equation}
In the deuteron lab system we have 
$\vec{p}_{N^*}\!=\!\vec{k}\!-\!\vec{p}_N$. 
Taking plane waves for all three outgoing particles, the expression
(\ref{41}) can be reduced to the form 
\begin{equation}\label{45}
M_{Sm_ST,\lambda m_d}^{IA}=\sum_{m_S^{\prime}}\langle Sm_S\,|\bigg[
\,t_{\gamma\eta}^{(T)}(W_{\gamma N_1})
\phi_{m_S^{\prime}m_d}(\vec{p}_2) 
-(-1)^{S+T}t_{\gamma\eta}^{(T)}(W_{\gamma N_2})
\phi_{m_S^{\prime}m_d}(\vec{p}_1)\,\bigg]|\,\,m_d\rangle\,,
\end{equation}
with upper index $(T)$ referring to the isoscalar $(s)$ or the 
isovector $(v)$ amplitude (\ref{25}) for $T=0$ and $T=1$, respectively.
The energy $W_{\gamma N_i}$ is defined in (\ref{44}) with
$\vec{p}_N=\vec{p}_{2(1)}$ for i=1(2).
The amplitude (\ref{45}) 
corresponds to the simple spectator-nucleon model.
The elementary operator $t^{(s/v)}_{\gamma\eta}$
produces the ($\gamma,\eta$) process on the 
nucleon, which takes the whole photon energy available after meson 
production.  Since in our approach the spectator nucleon is taken 
to be on-shell,
the active nucleon is off-shell as determined by the 
energy and momentum conservation at the corresponding vertex
\begin{eqnarray}\label{46}
\vec{p}&=&\vec{q}+\vec{p}_i-\vec{k} \,, \\
E_{\vec{p}}&=&\omega_{\vec{q}}+E_{\vec{p}_i}-\omega\,\,\,\,(i=1,2)\,.
\end{eqnarray}
In order to qualitatively explain the approximations concerning the 
rescattering terms, discussed below, we would like to demonstrate here 
some features of the reaction amplitude keeping in (\ref{37}) only the  
impulse approximation.
Due to the large transferred momentum associated with the large $\eta$ mass,
the $\eta$-angular distribution shows a fast decrease with
increasing angle, so that the major part of the cross section is 
concentrated in the forward direction. Therefore, we choose for the
discussion the forward angle region. The differential cross section
$d\sigma/d\Omega_{\vec{q}}$ 
calculated within the spectator model is shown in 
Fig.\ \ref{fig3} as a function of the lab photon energy, where in addition
the separate contribution from the various important partial waves
in the initial and final states are shown.  
Several conclusions can be drawn:

(i) Because of the large velocity of both outgoing nucleons near threshold, 
the high momentum components of the deuteron wave function become significant. 
As a consequence, a few MeV above threshold a rather large contribution
from the deuteron $D$-wave arises leading to a strong distructive
interference between $^3D_1$ and
$^3S_1$ states which reduces sizeably the pure $S$-wave cross section.
With increasing photon energy, the role of the $D$-wave 
decreases rapidly, e.g., 
at $\omega=650$ MeV this reduction  amounts to only 8$\%$.

(ii) The separate contributions
from the $^1S_0$ and $^3S_1$ states of the final 
$NN$-system clearly show the dominance of the singlet over the triplet 
$s$-wave.
The strong suppression of the $^3S_1$ state is due to the small 
isoscalar part of the $\eta$-photoproduction amplitude (in our calculation 
$t_{\gamma\eta}^{(s)}\!\approx$ 0.12\,$t_{\gamma\eta}^{(v)}$) so that 
the transition from the deuteron bound state ($T=0$,\,$S=1$) to the 
continuum $NN$-state ($T=1$,\,$S=0$) is most probable.
If one neglects the $D$-wave of the deuteron, the following approximate
relation holds
\begin{equation}\label{48}
\frac{d\sigma(^1S_0)}{d\sigma(^3S_1)}=\left|\frac{t_{\gamma\eta}^{(v)}}
{t_{\gamma\eta}^{(s)}}\right|^2
\frac{\sum_{m_d}|\langle S\!=\!0|\,\vec{\sigma}\!\cdot\!\vec{\varepsilon}\,|
\,m_d\rangle |^2}{\sum_{m_d}|\langle S\!=\!1|
\,\vec{\sigma}\!\cdot\!\vec{\varepsilon}\,|\,m_d\rangle|^2}
=\frac{1}{2}\left|\frac{t_{\gamma\eta}^{(v)}}{t_{\gamma\eta}^{(s)}}\right|^2
\approx 33\,.
\end{equation}
This strong dominance of the singlet $s$-wave state is slightly 
reduced by the presence of the $D$-component, in particular
near threshold. 

\subsection{The $NN$-rescattering}

\begin{figure}
\centerline{\psfig{figure=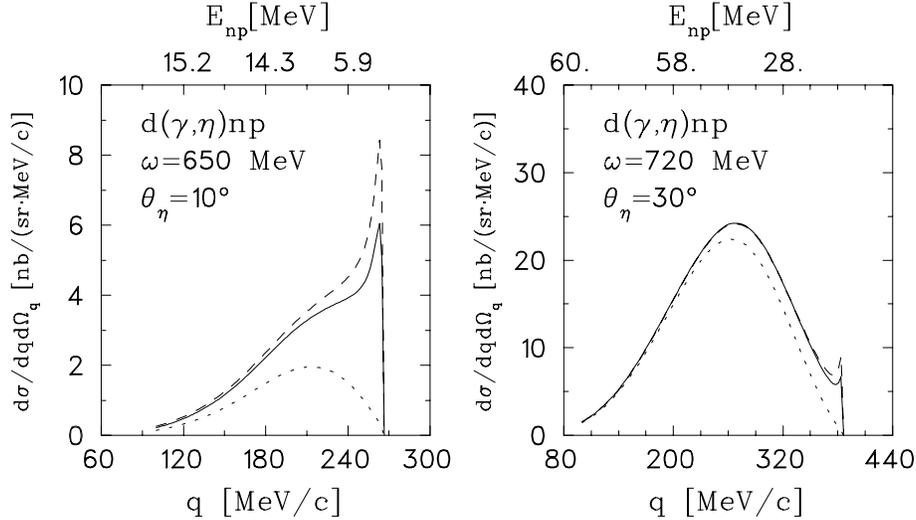,width=12cm,angle=0}}
\vspace*{.5cm}
\caption{
The $\eta$-meson spectra at forward emission angles  
in the $d(\gamma,\eta)np$ reaction for two different photon energies
and angles.
The dotted curves show the pure
impulse approximation, whereas the full curves include 
the interaction between the outgoing nucleons. 
The dashed curves represent the results obtained without 
the $D$-wave contribution to the 
$NN$-rescattering amplitude.      
The excitation energy $E_{np}$ in the final $NN$-system is indicated at the  
top abscissa.
}
\label{fig4}
\end{figure}

The diagram (b) in Fig.\ \ref{fig2} describes the $NN$-rescattering 
whose amplitude is given by
\begin{equation}\label{51}
M_{Sm_ST,\lambda m_d}^{NN}=
-\langle\vec{q},\vec{p}_1\vec{p}_2,
Sm_ST\,|\,T_{NN}G_{NN}V_{\eta N^*}^{\dagger}\,G_{N^*N}
(W_{\gamma N})\,\vec{\varepsilon}_{\lambda}\cdot\vec{J}_{N^*N}|\,m_d\rangle\,.
\end{equation}
The free nonrelativistic propagator of the two nucleons in terms of the 
relative $NN$-momenta $\vec{p}\!=\!(\vec{p}_2\!-\!\vec{p}_1)/2$ and 
$\vec{p}^{\,\prime}\!=\!(\vec{p}_2^{\,\prime}\!-\!\vec{p}_1^{\,\prime})/2$
has the form
\begin{equation}\label{53}
\langle\vec{p}\,|G_{NN}|\vec{p}^{\,\prime}\rangle
=\frac{M_N}{p^2-p^{\,\prime\,2}+i\epsilon}\,.
\end{equation}
The $NN$ dynamics in the final state is determined by the half-off-shell
nucleon-nucleon\\
$t$-matrix $T_{NN}$, which we evaluate for the same OBEPQ 
Bonn potential 
\cite{Mach87} as for the deuteron wave function. 
We restrict the $NN$-rescattering to the $^1S_0$ state. 
This approximation is justified by the previous observation 
that due to the spin-isospin selection rules the contribution
from the triplet $^3S_1$ state is at least one order of magnitude smaller
than that from the $^1S_0$ state. As for the higher partial waves,
their distortion is expected to be of minor importance compared to the 
resonant $^1S_0$-wave, since in the 
kinematical region considered in this work the relative kinetic energy 
in the $NN$-system is less than 72 MeV (see also 
\cite{Laget78}). 
Within this approximation, a straightforward calculation yields 
finally 
\begin{equation}\label{55}
M_{001,\lambda m_d}^{NN}=
M_N\sum_{m_S^{\prime}}\langle 00\,|\,\int\frac{2t_{\gamma\eta}^{(v)}
(W_{\gamma N})\,T_{NN}(p^{\,\prime},p)}{p^2-p^{\,\prime\,2}+i\epsilon}
\phi_{m_S^{\prime}m_d}(\vec{p}_N)\frac{d^3p^{\,\prime}}{(2\pi)^3}
|\,\,m_d\rangle\,,
\end{equation}
with $\vec{p}_N=\frac{\vec{k}-\vec{q}}{2}+\vec{p}^{\,\prime}$.
Since, as was noticed above, the deuteron $D$-wave component plays a
non-negligible role in the IA,
it was also taken into account in the matrix element (\ref{55}).
The influence of $NN$-rescattering is demonstrated in Fig.\ \ref{fig4}
for the $\eta$-momentum distribution at a given angle 
$\theta_{\eta}$. 
As expected, the strong interaction between the final 
nucleons in the $^1S_0$ state changes drastically the cross section for 
large $\eta$ momentum values. When $q$ reaches its maximum, the excitation  
energy $E_{np}$ 
in the $np$-pair vanishes, and thus the resonant 
$^1S_0$ state appears as a rather narrow peak. 
The same effect is noted for charged $\pi$-photoproduction on the deuteron
\cite{Laget78}
as well as for deuteron electrodisintegration \cite{FA79}. 
In principle, the experimental observation of this peak in the high
$\eta$-momentum spectrum may serve as another evidence for the  
isovector nature of the $N^*$ photoexcitation. In the hypothetical 
case $t_{\gamma\eta}^{(s)}\gg t_{\gamma\eta}^{(v)}$, the low-energy 
$NN$-rescattering would be dominated by the $^3S_1$ state,
which does not exhibit any resonant behaviour at $E_{np}\!\approx$ 0.
 
In conclusion, one sees that the role of $NN$-rescattering is quite  
important, especially for small photon energies.
At higher energies, the main part of the cross section
is dominated by the IA, which gives a rather broad quasifree bump,
where the role of   
$NN$-rescattering is expected to be of minor importance.
The contribution from the deuteron $D$-wave is only significant 
close to threshold. Its effect is a
destructive interference between the $^3D_1\!\rightarrow\!^1S_0$
and the dominant $^3S_1\!\rightarrow\!^1S_0$
transitions (see Fig.\ \ref{fig4}).

\begin{figure}
\centerline{\psfig{figure=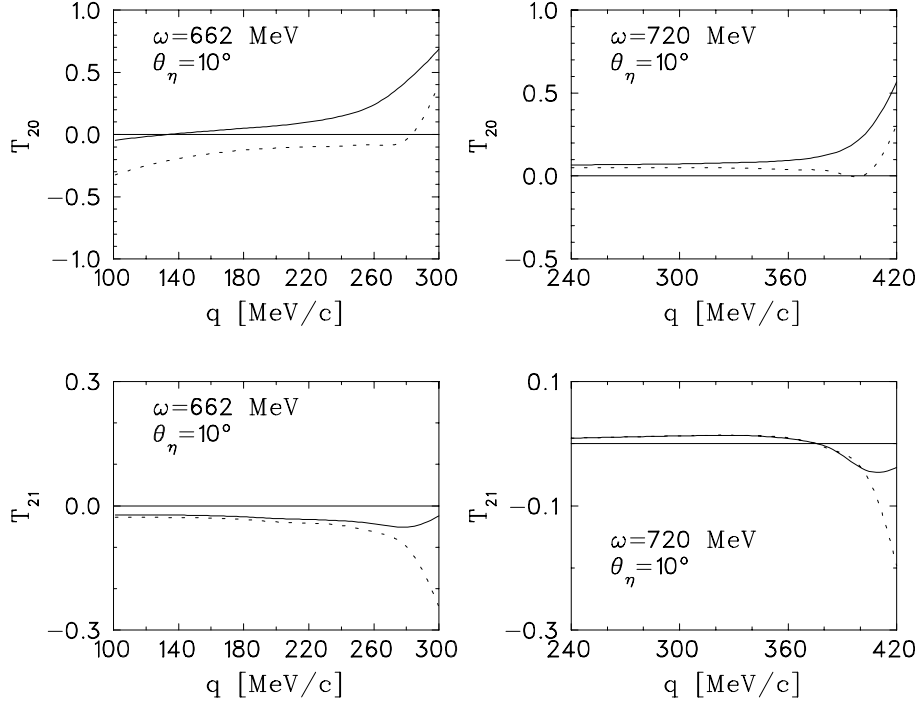,width=12cm,angle=0}}
\vspace*{.5cm}
\caption{
Influence of $NN$-rescattering on several tensor 
target asymmetries for two photon lab energies as function of the 
$\eta$-momentum.
Notation of the curves as in Fig.\ \protect{\ref{fig4}}.
}
\label{fig5}
\end{figure}

In Fig.\ \ref{fig5} we depict also the tensor target asymmetries
for $\vec{d}(\gamma,\eta)np$ at two different photon 
energies. One readily notes that the asymmetry $T_{20}$ is strongly
affected by $NN$-rescattering.
The reason for this is that $T_{20}$
is directly defined by the difference between the cross section on the 
deuteron with spin parallel and perpendicular to the photon beam
direction (see (\ref{T20})).
Because of the spin-transverse form of the elementary operator (\ref{25}), 
the transition $^3S_1\!\rightarrow\!^1S_0$ may occur 
in the $\eta$-photoproduction 
process with only opposite directions of photon and deuteron 
spin. Consequently, the strong $^1S_0$ interaction between the 
final nucleons,  
enhancing the probability of this transition, increases the contribution
from the matrix elements $M_{Sm_ST,\lambda m_d}$ in (\ref{T20})
with $m_d=\pm$1.
The other target asymmetries,  
$T_{22}$ and $T_{11}$ which are not shown here, 
appear not to be sensitive to the 
$NN$-rescattering contribution. For the photon asymmetry $\Sigma$ we have
a trivial relation $\Sigma=0$, which follows from our neglect
of other terms besides the $N^*$-excitation amplitude in the elementary
operator.

\subsection{The $\eta N$-rescattering}

The $\eta N$-rescattering amplitude given by the the diagram (c) 
in Fig.\ \ref{fig2} reads
\begin{equation}\label{61}
M_{Sm_ST,\lambda m_d}^{\eta N}=
-\langle\vec{q},\vec{p}_1\vec{p}_2,
Sm_ST\,|\,\frac{M_N}{2\mu W_{\eta N}}t_{\eta N}G_{\eta N}
V_{\eta N^*}^{\dagger}\,G_{N^*N}(W_{\gamma N})\,\vec{\varepsilon}_{\lambda}
\cdot\vec{J}_{N^*N}|\,m_d\rangle\,,
\end{equation}
where $\mu\!=\!\frac{M_Nm_{\eta}}{M_N+m_{\eta}}$.
Similarly to the $NN$-case (\ref{53}), we use the nonrelativistic 
propagator for the relative $\eta N$-motion
\begin{equation}\label{62}
\langle\vec{q}_{\eta N}|G_{\eta N}|\vec{q}_{\eta N}^{\,\prime}\rangle
=\frac{2\mu}{q_{\eta N}^2-
q_{\eta N}^{\,\prime\,2}+i\epsilon}\,.
\end{equation}
In the spirit of our isobar approach, we consider the $\eta N$-interaction
to proceed only via the $N^*$ formation, which gives  
the $t$-matrix in the form 
\begin{equation}\label{63}
t_{\eta N}(W_{\eta N})=v_{N^*}^{\dagger}g_{N^*}(W_{\eta N})v_{N^*}=
\frac{g_{\eta NN^*}^2}{W_{\eta N}-M_{N^*}+\frac{i}{2}\Gamma(W_{\eta N})}\,,
\end{equation}
where $W_{\eta N}$ is the $\eta N$ invariant mass and other parameters 
are defined in Section \ref{section2}. Collecting the 
various pieces, we obtain for the amplitude (\ref{61}) 
\begin{eqnarray}\label{64}
M_{Sm_ST,\lambda m_d}^{\eta N}&=&\sum_{m_S^{\prime}}
\langle Sm_S\,|\,\int\Bigg[\frac{M_N}{W_{\eta N_2}}\,
\frac{t_{\gamma\eta}^{(T)}(W_{\gamma N_1})
t_{\eta N}(W_{\eta N_2})}{q_{\eta N_2}^2-q_{\eta N}^{\,\prime\,2}+i\epsilon}
\phi_{m_S^{\prime}m_d}(\vec{p}_{N_2}) \\ \nonumber
&-&(-1)^{S+T}\frac{M_N}{W_{\eta N_1}}\,
\frac{t_{\gamma\eta}^{(T)}(W_{\gamma N_2})
t_{\eta N}(W_{\eta N_1})}{q_{\eta N_1}^2-q_{\eta N}^{\,\prime\,2}+i\epsilon}
\phi_{m_S^{\prime}m_d}(\vec{p}_{N_1})\Bigg]
\frac{d^3q_{\eta N}^{\,\prime}}{(2\pi)^3}
|\,\,m_d\rangle\,,
\end{eqnarray}
with $\,\vec{p}_{N_i}\!=\!\frac{\mu}{m_{\eta}}(\vec{k}-\vec{p}_i)
+\vec{q}_{\eta N}^{\,\prime}\,$ and
$\,\vec{q}_{\eta N_i}\!=\!\frac{m_{\eta}\vec{p}_i-
M_N\vec{q}}{M_N+m_{\eta}}$ $(i=1,2)\,$.

\begin{figure}
\centerline{\psfig{figure=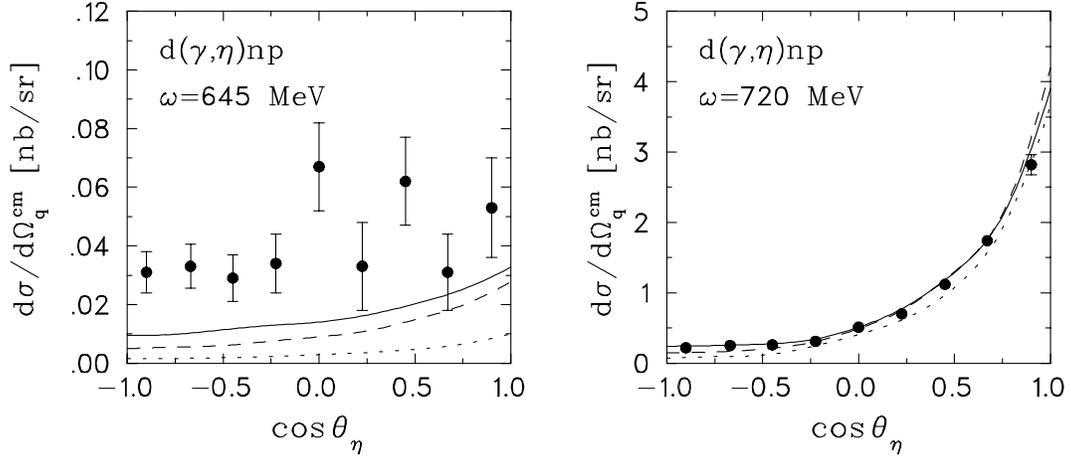,width=14cm,angle=0}}
\vspace*{.5cm}
\caption{
The differential $d(\gamma,\eta)np$ cross section, calculated 
in the $\gamma d$ c.m.\ frame. Shown are the IA
prediction (dotted lines), the successive addition of $NN$ (dashed
lines) and $\eta N$ (full lines) rescattering. 
The experimental points represent the inclusive $\gamma d\!\rightarrow\eta X$
measurements from Ref.\ \protect{\cite{Kru95}}.
}
\label{fig6}
\end{figure}

The effect of $\eta N$-rescattering is demonstrated in Fig.\ \ref{fig6}
for the $\eta$-angular distribution in the $\gamma d$ c.m.\ frame,
where we also compare our results with experimental data for the
inclusive reaction $\gamma d\!\rightarrow\!\eta X$ \cite{Kru95}.
In view of the small isoscalar part $t_{\gamma\eta}^{(s)}$ of the
elementary amplitude and the large momentum mismatch, the contribution from 
the coherent $\gamma d\!\rightarrow\!\eta d$ process is expected to be
negligible and, thus, the inclusive $\eta$-spectrum is dominated by the 
deuteron break-up channel. In the backward direction,
the increase from  
$\eta N$-rescattering is in part kinematically enhanced.
At these angles, the nucleons leave the interaction region  
with large momenta. Therefore, the spectator model gives a very 
small cross section understimating the 
data by roughly a factor of \,3\, for $\omega=720$ MeV. 
In this situation, the 
$\eta N$-rescattering mechanism, allowing the large transferred momentum 
to be shared between two participating nucleons, becomes much more effective. 
The resonant character of the $\eta N$-interaction appears more pronounced 
at forward angles. Although its strong inelasticity decreases the cross section
at high photon energies, close to threshold this effect is more than
compensated by the attraction in the $\eta N$-system.

\begin{figure}
\centerline{\psfig{figure=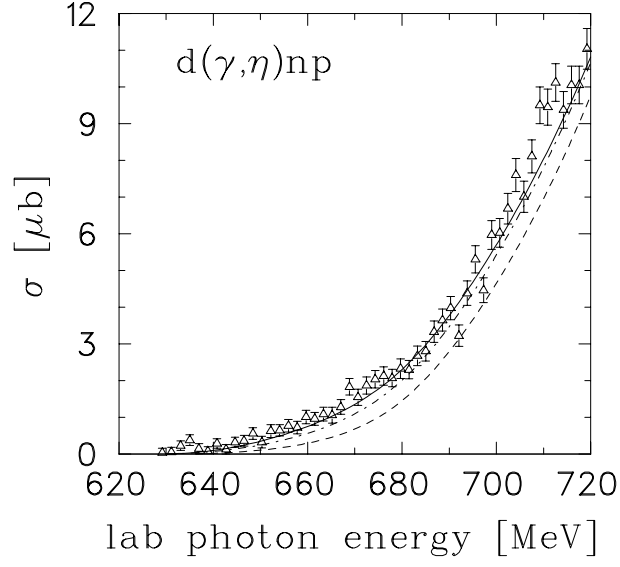,width=8cm,angle=0}}
\vspace*{.5cm}
\caption{
Results of our calculations for the total 
$\gamma d\!\rightarrow\eta np$ cross section compared with inclusive
$\gamma d\!\rightarrow\eta X$ experimental data \protect{\cite{Kru95}}.
The full and dashed lines represent the results obtained with and without
allowance for rescattering of the final particles, respectively. 
The dash-dotted line includes only IA and $NN$-rescattering.
}
\label{fig7}
\end{figure}

Finally, we show in Fig.\ \ref{fig7} the total 
$\gamma d\!\rightarrow\!\eta np$ cross section. 
One readily notes,
that the simple spectator approach cannot 
describe the experimental data close 
to the threshold (see also \cite{Sau95}). 
As has been discussed in the Introduction, 
at small photon energies the spectator model considers the 
$\gamma d\!\rightarrow\!\eta np$ process to proceed mainly through
the high deuteron Fourier component, which appears 
only with small probability. The final state interaction provides 
a mechanism for bypassing this suppression. 
Our model predicts quite a significant contribution 
from $NN$-rescattering. It turns out to be dominant  
in the vicinity of the threshold and still increases the spectator 
results 
by about 10$\%$ at $\omega=720$ MeV. The $\eta N$-interaction is 
relatively less important, but is also significant
mainly through the constructive interference between $NN$- and 
$\eta N$-rescattering contributions.
With inclusion of the rescattering effects, we are able to reproduce
the experimental cross section with $\alpha=0.11$.    

\begin{figure}
\centerline{\psfig{figure=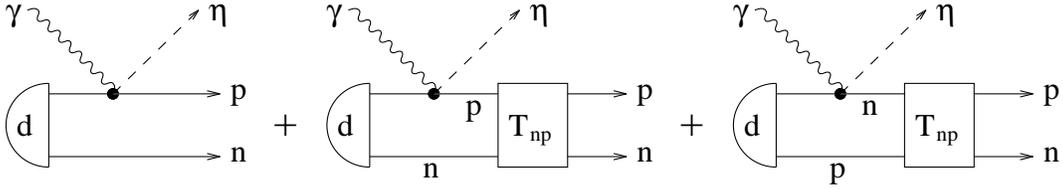,width=14cm,angle=0}}
\vspace*{.5cm}
\caption{
Different mechanisms of the $\gamma d\rightarrow\eta np$ process
which lead to the same $np$-configuration in the final state. 
}
\label{fig8}
\end{figure}

At this point, we would like to discuss briefly the possibility
of extracting the ratio  
$|t^{(n)}_{\gamma\eta}/t^{(p)}_{\gamma\eta}|$
from the corresponding ratio of exclusive $d(\gamma,\eta n)p$
and  $d(\gamma,\eta p)n$ yields as has been done in Ref.\ \cite{Hoff97}.
Strictly speaking, the simple relation
\begin{equation}\label{71}
\left|\frac{t^{(n)}_{\gamma\eta}}
{t^{(p)}_{\gamma\eta}}\right|=
\sqrt{\frac{d\sigma_{d(\gamma,\eta n)p}}{d\sigma_{d(\gamma,\eta p)n}}}\,,
\end{equation}
consistent with a pure spectator approach is not valid away from
quasifree kinematics, in particular, when final 
state interaction is taken into account. This is due to the fact, that 
rescattering ``mixes'' the single proton and neutron $\eta$-photoproduction
processes, as is illustrated schematically in Fig.\ \ref{fig8},
taking as an example the $NN$-interaction term. 
Thus, in the region where the rescattering effects are dominant, 
the right hand side of (\ref{71}) approximates unity in contrast to the 
elementary ratio.

\begin{figure}
\centerline{\psfig{figure=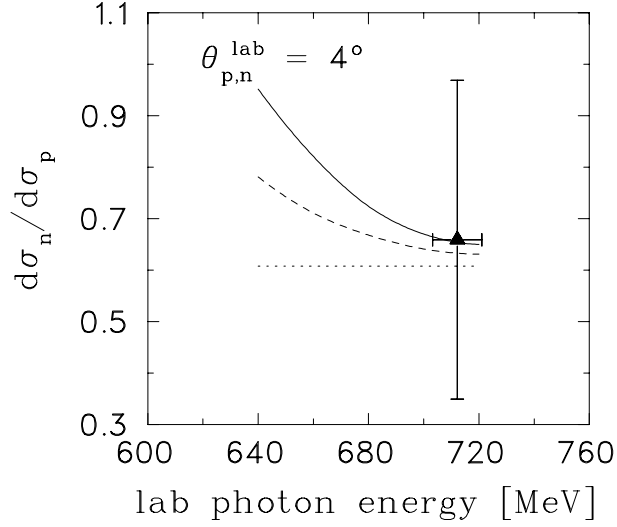,width=8cm,angle=0}}
\vspace*{.5cm}
\caption{
Ratio of neutron to proton angular distribution
calculated for the reaction $d(\gamma,\eta N)N$ as a function of incident 
photon energy.   
The dotted curve is 
the free nucleon $\sigma_n/\sigma_p$ ratio. The dashed curve
shows the spectator model calculation, while 
$NN$- and $\eta N$-rescatterings are taken into account in the full curve.
The data point is taken from the $\eta N$ coincidence  measurements of 
Ref.\ \protect{\cite{Hoff97}}. 
}
\label{fig9}
\end{figure}

In general, we can conclude from the above investigations, that the relation  
(\ref{71}) works well for photon energies not too close to the threshold.
As a confirmation, we have plotted in Fig.\ \ref{fig9}
the theoretical ratio  
$\frac{d\sigma}{d\Omega_n}/\frac{d\sigma}{d\Omega_p}$ of the 
differential cross sections for the reaction 
$d(\gamma,\eta N)N$ without and with rescattering versus the
photon lab energy $\omega$ and compare it to the corresponding value
for free nucleons. The difference between the ratio  
for free nucleons and that for the deuteron, obtained within
the spectator model, is due to the fact, that to the integration
over nucleon momenta not only fast moving active nucleons but also 
slow ones partly contribute.  
One can see, that the additional final state interaction encreases 
even more the ratio when approaching the threshold.
However, for the energies considered
in \cite{Hoff97}, its role is negligible, and from this point of view
the validity of (\ref{71}) is fully justified. The ratio of the neutron to 
the proton cross sections $\sigma_n/\sigma_p=0.61$, 
corresponding to $\alpha=0.11$, which we have used in this work and 
which reproduces well the  
$\gamma d\!\rightarrow\!\eta X$ data of \cite{Kru95}, 
slightly understimates the one presented in Ref.\ \cite{Hoff97}
$\sigma_n/\sigma_p=0.68\pm0.06$.

\begin{figure}
\centerline{\psfig{figure=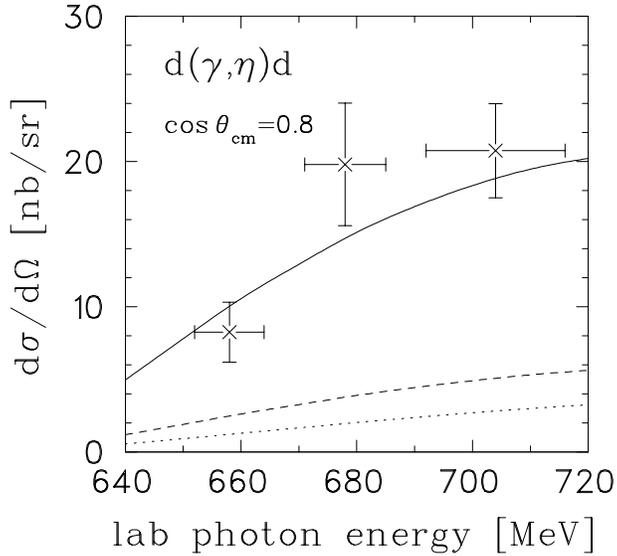,width=8cm,angle=0}}
\vspace*{.5cm}
\caption{
Differential $d(\gamma,\eta)d$ cross section calculated with different 
values of the parameter $\alpha=t^{(s)}_{\gamma\eta}/t^{(p)}_{\gamma\eta}$.
The dotted, dashed, and full curves correspond 
to $\alpha=0.09$, 0.13 and 0.26, respectively. 
The data are taken from Ref.\ \protect{\cite{Hoff97}}.
}
\label{fig10}
\end{figure}

Finally, we would like to make a short comment on the coherent photoproduction
process on the deuteron.
We first note, that the sensitivity of the break-up 
cross section to a variation of $\alpha$ is much less pronounced than 
for the coherent reaction, which
is proportional to $|\,t_{\gamma\eta}^{(s)}|^2$. 
The present study within the isobar model has shown, that the experimental 
uncertainity in the $\gamma d\!\rightarrow\!\eta np$ total cross section
for photon energies between threshold and
720 MeV \cite{Kru95} fixes the parameter $\alpha$
within the limits of $\alpha=0.11\pm0.02$. In Fig.\ \ref{fig10} we 
demonstrate the strong sensitivity on $\alpha$ 
of the $\eta$-angular distribution
of the coherent reaction, which we have calculated 
within the conventional impulse approximation. In addition to the
$S_{11}$(1535) resonance, the contributions from $D_{13}$(1520) 
as well as the nucleon pole term and $\omega$-meson exchange in the t-channel
were included in the elementary operator (for further detailes see 
\cite{Fix97a}). We do not touch here upon the question about the role 
of meson rescattering in the coherent reaction,
which was investigated
in \cite{Kamalov97,Hoshi79,Hald89}. However, one can expect,
that due to its two-particle character, this effect
will not strongly affect the coherent process 
in the region of forward $\eta$-angles which is associated with relatively
small momentum transfer. As can be seen from Fig.\ \ref{fig10} even the
largest permissible value $\alpha=0.13$ gives a cross section, which is
by more than a factor three below the data. Only with a rather large  
isoscalar amplitude corresponding to $\alpha=0.26$ the impulse approximation
could describe the data well. This seeming inconsistency between the 
coherent and incoherent reaction with respect to the precise value 
of the parameter $\alpha$ certainly needs further experimental and 
theoretical studies.
  
\section{Conclusion}

We have studied incoherent $\eta$-photoproduction on the deuteron 
in the near threshold region. 
The elementary photoproduction amplitude, based on a pure isobar model,
for which the $\gamma NN^*$ and $\eta NN^*$ vertices
have been fitted, gives a good description of the experimental 
$p(\gamma,\eta)p$ cross section data. 
As for the deuteron amplitude, we have considered in addition to the 
spectator-nucleon approach also $NN$-rescattering in the
most important $^1S_0$ channel 
and $\eta N$-rescattering. While the pure spectator model 
produces only a small fraction of the observed cross section near threshold, 
the strong attractive $NN$-interaction in the resonant $^1S_0$ state 
enhances it considerably. The role of $\eta N$-rescattering
is found to be less important. With inclusion of both rescattering effects 
we obtain quite a satisfactory agreement with the available data on the 
$\gamma d\!\rightarrow\!\eta X$ differential and total cross sections,
using as input for the ratio of  
isoscalar to proton amplitude $\alpha=0.11$. 
This value is roughly consistent with $\alpha=0.09\pm0.02$, 
extracted from the recently measured ratio 
$d\sigma_{d(\gamma,\eta n)p}/d\sigma_{d(\gamma,\eta p)n}$.
Furthermore, we have shown, 
that this extraction can be justified within the frame 
of the pure spectator-nucleon model, since the role of final state
interaction under the corresponding kinematical conditions is negligible.

\acknowledgments
A.\ F.\ is grateful to B.\ Krusche, R.\ Schmidt, 
M.\ Schwamb and P.\ Wilhelm for fruitful discussions. 
He would also like to thank 
the theory group of Professor H.\ Arenh\"ovel as well as 
Professor H.\ Str\"oher, Monika Baumbusch 
and many scientists of the Institut f\"ur Kernphysik
at the Mainz University for the very kind hospitality.

\vspace{3cm}

\end{document}